# Realization of Topology-controlled Photonic Cavities in a Valley Photonic Crystal

Bei Yan[1,2,#], Baoliang Liao[3,#], Fulong Shi[4,#], Xiang Xi[5,#], Yuan Cao[3], Kexin Xiang[1], Yan Meng[5], Linyun Yang[6], Zhenxiao Zhu[1], Jingming Chen[1], Xiao-Dong Chen[4,*], Gui-Geng Liu[7,†], Baile Zhang[7,‡], Zhen Gao[1,§]

[1]State Key Laboratory of Optical Fiber and Cable Manufacturing Technology, Department of Electronic and Electrical Engineering, Guangdong Key Laboratory of Integrated Optoelectronics Intellisense, Southern University of Science and Technology, Shenzhen 518055, China

[2]Hubei Province Key Laboratory of Systems Science in Metallurgical Process, and College of Science, Wuhan University of Science and Technology, Wuhan 430081, China

[3]Guangdong Province Key Laboratory of Optical Fiber Sensing and Communications, Institute of Photonics Technology, Jinan University, Guangzhou 510632, China

[4]School of Physics & State Key Laboratory of Optoelectronic Materials and Technologies, Sun Yat-sen University, Guangzhou 510275, China

[5]School of Electrical Engineering and Intelligentization, Dongguan University of Technology, Dongguan, 523808, China

[6]College of Aerospace Engineering, Chongqing University, Chongqing, 400030, China

[7]Division of Physics and Applied Physics, School of Physical and Mathematical Sciences, Nanyang Technological University, 21 Nanyang Link, Singapore 637371, Singapore

We report an experimental realization of a new type of topology-controlled photonic cavities in valley photonic crystals by adopting judiciously oriented mirrors to localize the valley-polarized edge states along their propagation path. By using microwave frequency- and time-domain measurements, we directly observe the strong confinement of electromagnetic energy at the mirror surface due to the extended time delay required for the valley index flipping. Moreover, we experimentally demonstrate that both the degree of energy localization and quality factors of the topology-controlled photonic cavities are determined by the valley-flipping time which is controlled by the topology of the mirror. These results extend and complement the current design paradigm of topological photonic cavities.

Photonic cavities that efficiently confine light within small volumes for extended durations play a pivotal role in modern photonics with wide applications ranging from low-threshold lasers [1,2], ultra-small filters [3,4], photonic chips [5], quantum information processing [6,7] to optical communications [8,9]. These photonic cavities typically include Fabry-Pérot cavities with two parallel mirrors, whispering-gallery-mode cavities [10,11] formed by total internal reflection on closed concave surfaces, and photonic crystal cavities that employ photonic bandgaps to trap light at point or line defects [2,12,13].

Recently, topological photonic cavities [14-20] with robust topological protection have attracted great attention and shown promising applications in topological lasers [21-23], integrated photonics [24,25], robust optical delay lines [26], and quantum optics [27,28]. Among these, a novel type of topology-controlled photonic cavities based on the near-conservation of photonic valley degree of freedom (DOF) was theoretically proposed by placing a judiciously oriented mirror at the termination of a valley photonic crystal (VPC) waveguide to localize the valley-polarized edge states [29]. These robust valley edge states, rooted in the nonzero Berry curvatures within the Brillouin zone [30-34], require the flipping of their valley-locking momentum for reflection. Remarkably, this momentum flip involves an extended time delay controlled by the geometry of the terminating mirror. Therefore, when the effective reflection time $t_r$ (extended time delay) is sufficiently long, the electromagnetic energy becomes tightly confined at the mirror's surface, forming a subwavelength topology-controlled photonic cavity. However, due to the challenges associated with measuring the electromagnetic field distributions and the extended time delay, experimental realization of this novel topology-controlled photonic cavity remains elusive.

In this Letter, we report the first experimental realization of a topology-controlled photonic cavity based on the near-conservation of the valley DOF in a VPC. We construct this topology-controlled photonic cavity by terminating a VPC waveguide with a metallic mirror whose geometry and orientation dictate the inter-valley flipping rate and reflection time delay. By employing microwave near-field mapping and pulse transmission measurements, we directly observe the enhanced electric field distributions and extended time delay of the topology-controlled photonic cavity. Furthermore, we experimentally demonstrate that different orientations and shapes of the mirrors result in varying time delays required for the valley index flipping, leading to distinct levels of energy confinement and quality factors. These results expand the research scope of topological photonics and enrich the fundamental physical principles of photonic cavities.

We begin with a VPC consisting of a triangular lattice of metallic tripods suspended between two parallel metallic plates, as illustrated in Figs. 1(a)-(c). In this study, we only focus on the transverse magnetic (TM) modes with electric fields polarized along the $z$-axis. As shown in Fig. 1(d), the simulated bulk band structure of the VPCs exhibits a photonic bandgap ranging from 5.8 to 6.2 GHz (gray region). Then we construct a VPC waveguide containing a domain wall separating two topological nontrivial VPCs with oppositely oriented metallic tripods and opposite Berry curvatures [35,36], as shown in the left panel of Fig. 1(e) and the right panel of Fig. 1(d). Therefore, the two VPCs have opposite valley-Chern indices and their domain wall supports $K(K')$ valley-polarized edge states propagating forward (backward), as shown in the right panel of Fig. 1(e), the eigenmode A of the valley edge state is tightly localized and propagates along the domain wall.

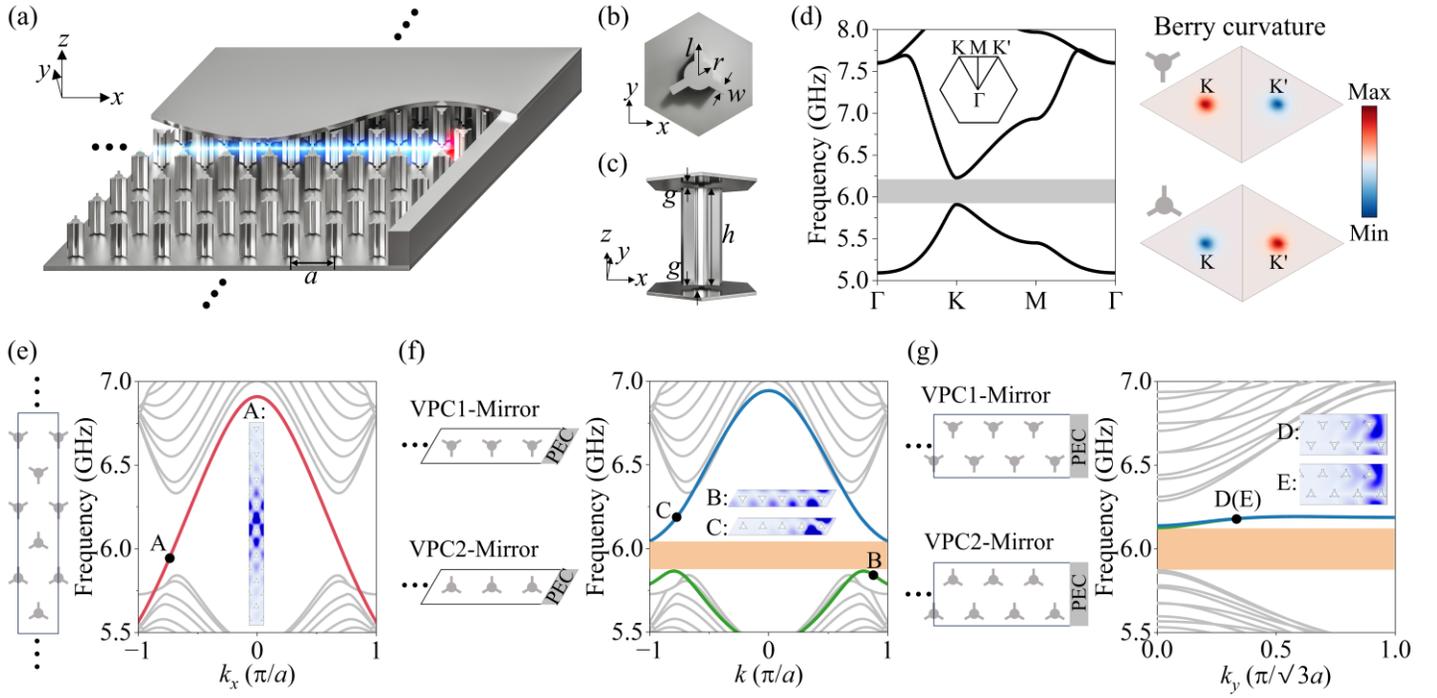

FIG. 1. Design of a topology-controlled photonic cavity by terminating a VPC waveguide with a metallic mirror. (a) Schematic of the topology-controlled photonic cavity consisting of two VPCs with opposite valley Chern numbers and a metallic mirror placed at the end of the VPC waveguide. (b) Top and (c) side views of a unit cell of the VPC. A metallic tripod is symmetrically sandwiched between two parallel metallic plates with inner radius $r = 3.68$mm, arm length $l = 7.95$mm, arm width $w = 2.21$mm, tripod height $h = 34.6$mm, and two small air gaps $g = 1.1$mm. In the experiment, a perforated dielectric foam (ROHACELL 31HF) with a relative permittivity of 1.04 and a loss tangent of 0.0025 is adopted to fill the air gap. The lattice constant is $a = 36.8$mm. (d) Simulated bulk band structure (left panel) and Berry curvature (right panel) of the lower band of the VPC. The inset shows the first Brillouin zone. (e) Supercell of the VPC waveguide formed by two VPCs with opposite valley Chern numbers and its simulated dispersion relationship. (f) Supercells of two interfaces between two VPCs and zigzag PEC termination and their dispersion relationships. (g) Supercells of two interfaces between two VPCs and armchair PEC termination and their dispersion relationships. Insets show the electric field distributions of the eigenmodes A, B, C, D, and E.

To realize a topology-controlled photonic cavity, we terminate the VPC waveguide with a perfect electric conductor (PEC) mirror, as schematically shown in Fig. 1(a) and the left panels of Figs. 1(f)-1(g). When a rightward-propagating valley edge state encounters the metallic mirror, it has three different scattering channels, the upper VPC1-mirror interface, the lower VPC2-mirror interface, and the VPC waveguide. We first study the upper and lower interfaces between the VPCs and the metallic termination. Depending on the orientation of the metallic mirror, we consider two types of terminations: zigzag [Fig. 1(f)] and armchair [Fig. 1(g)]. For the zigzag (armchair) termination we examine two supercells with a metallic mirror oriented $60^o$ ($90^o$) to the propagation direction of the $K$-valley polarized edge states and analyze their eigenfrequency spectra, as shown in Fig. 1(f) [Fig. 1(g)], the green (blue) line represents the dispersion of the trivial interface states supported by the upper VPC1-mirror interface (lower VPC2-mirror interface). The gapped trivial interface modes B and C (D and E) are localized near the mirror surface with a complete no-interface-mode bandgap [orange regions in Figs. 1(f)-1(g)], indicating that when a valley-polarized edge state within the no-interface-mode bandgap encounters the mirror termination, it cannot leak through the upper and lower VPC-mirror interface channels.

The last scattering channel is backscattering which requires the valley index of the topological edge states to reverse from $K$ to $K'$. Consequently, the valley-index-flipping rate is equivalent to the leaky rate of the topology-controlled photonic cavity. Interestingly, previous studies [30,32] have revealed that the valley-index-flipping rate depends on the topology of the perturbations and the zigzag termination (perturbation along the principal axes of the lattice) can significantly suppress the inter-valley scattering because of the valley conservation. In contrast, the situation is different for the armchair termination (perturbation along the orthogonal direction) which produces a much higher valley-index-flipping rate because of the valley conservation breaking. Therefore, if the VPC is terminated with a zigzag-oriented PEC mirror, the valley-flipping rate should be minimized, and the valley edge states will be tightly localized near the mirror surface, forming a subwavelength topology-controlled photonic cavity.

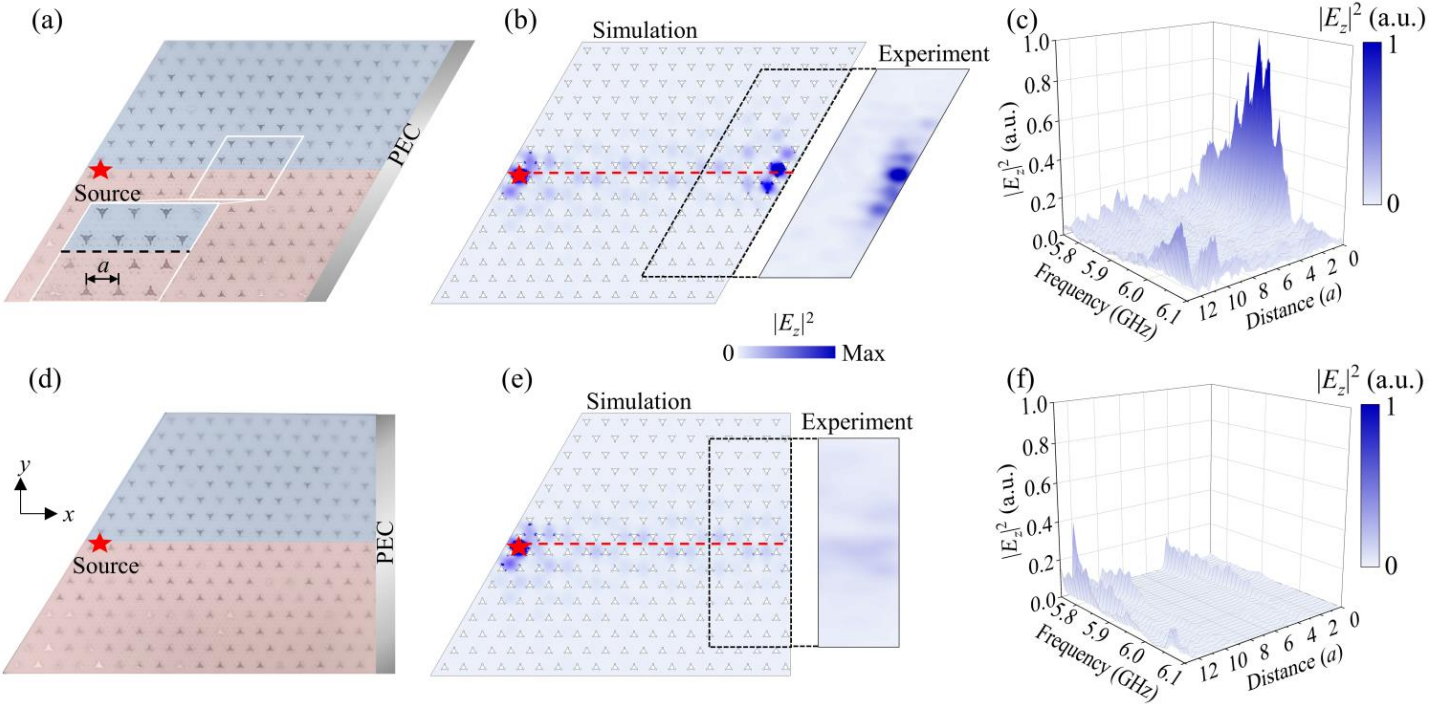

FIG.2. Observation of the frequency-domain electric field distributions of the topology-controlled photonic cavity. (a), (d) Top view of the zigzag and armchair topology-controlled photonic cavities. The top metallic plate is removed for clarity. (b), (e) Simulated and measured electric field distributions of the zigzag and armchair topology-controlled photonic cavities at 5.90 GHz. (c), (f) Measured electric field intensity along the domain wall (red dashed lines) of the zigzag and armchair topology-controlled photonic cavities with different frequencies.

Next, we experimentally demonstrate this novel topology-controlled photonic cavity. Fig. 2(a) and Fig. 2(d) show the fabricated samples with zigzag and armchair terminations, respectively, the upper metallic plate is removed to see the inner structures. A vertical dipole antenna (red star) was placed at the left end of the domain wall to excite the TM-polarized valley edge states and the topology-controlled photonic cavity. Another dipole antenna was inserted into the sample and scanned step-by-step to measure the $E_z$ field distributions. The source and probe antennas are connected to a vector network analyzer (Keysight E5080). For the topology-controlled photonic cavity with zigzag termination that preserves the valley conservation, as shown in Fig. 2(b), the simulated and measured $E_z$ field distributions exhibit large electric field enhancement near the right termination, indicating the zigzag topology-controlled photonic cavity has good field confinement and high-quality factor because of the minimized valley-index-flipping rate at the zigzag termination. For completeness, we plot the measured $E_z$ field amplitude along the domain wall (red dashed line) across a wide frequency range in Fig. 2(c), revealing that only the valley edge states within the no-interface-mode bandgap [orange regions in Figs. 1(f)-1(g)] can be tightly localized with significantly enhanced electric fields near the mirror surface. In contrast, for the topology-controlled photonic cavity with armchair termination, as shown in Figs. 2(e)-2(f), the simulated and measured $E_z$ fields are almost uniformly distributed in the VPC waveguide and have no obvious field localization and enhancement near the mirror surface within a wide frequency range (even in the no-interface-mode bandgap), indicating that the armchair topology-controlled photonic cavity with valley breaking has an ultralow quality factor because of its high valley-index-flipping rate, matching well with the theoretical prediction [29].

As concluded previously, the zigzag (armchair) termination minimizes (maximizes) the valley-flipping rate and delays the reflection of the valley edge states for a long (short) time. To experimentally verify this underlying physical mechanism of the topology-controlled photonic cavity, we perform time-domain pulse measurements to extract the reflection delay time at the mirror surface directly. As schematically shown in Fig. 3(a), we employ an arbitrary waveform generator (AWG, Tektronix AWG70002A) to generate a 100-MHz-wide Gaussian pulse centered around 5.9 GHz and a time interval of 100 ns between adjacent pulses. Then the Gaussian pulse was magnified by an electrical amplifier (EA, JCA910-4139PSTA) and launched into the VPC waveguide using a dipole antenna (red star). Simultaneously, another dipole antenna (blue star) was placed at the left side of the source antenna to capture the directly transmitted and reflected pulse signal using an oscilloscope (OSC, Teledyne LeCroy

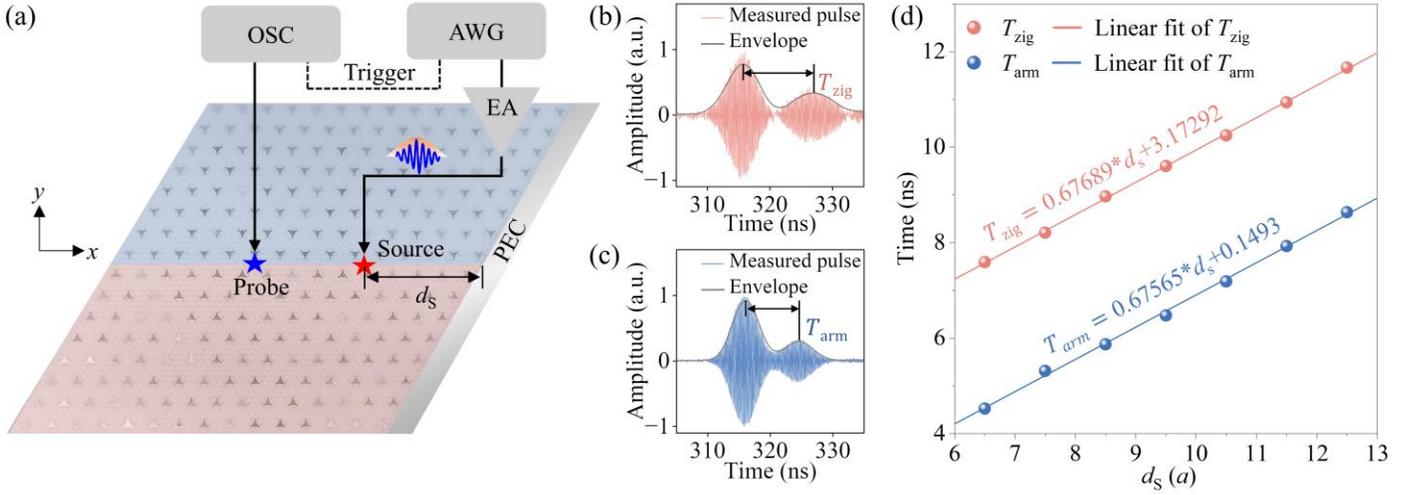

FIG.3. Observation of the time-domain pulse transmission in the zigzag and armchair topology-controlled photonic cavities. (a) Experimental set-up for the time-domain pulse measurements. The red and blue stars represent the source and probe antennas. The AWG is adopted to generate the input Gaussian pulse, the EA is used to magnify the pulse signal, and the OSC is exploited to capture the pulse signal. (b), (c) The measured time-domain pulse signals (normalized by their maximal amplitudes) of the zigzag and armchair topology-controlled photonic cavities. The gray lines represent the envelope of the measured pulse signals. (d) The measured time differences (spheres) between the directly transmitted and reflected pulse signals with different source positions and their linear fits (lines).

MCM-Zi, 10-36Zi). The measured pulse signals for the zigzag and armchair terminations are presented in Figs. 3(b) and 3(c), respectively. In each period we observe two pulse peaks. The first peak corresponds to the directly transmitted pulse signal excited by the source and the second peak corresponds to the reflected pulse signal from the right metallic mirror. The time difference $T$ between the two pulse peaks can be given by $T = 2d_s/v_g + t_r$, where $d_s$ is the distance between the pulse source (red star) and the metallic mirror, $v_g$ is the group velocity of the valley edge states, and $t_r$ is the delay time of the intervalley flipping. By varying the source positions, we measured the $d_s$-dependent time differences between the two pulse peaks for the zigzag (red spheres) and armchair (blue spheres) terminations, as shown in Fig. 3(d). Through linear fitting of the measured results, we extracted the group velocity $v_g$ of the valley edge state from the slope and the valley-index-flipping time $t_r$ from the intercept of the fitting line, respectively. The measured group velocities for the zigzag and armchair terminations are $0.362c$ and $0.363c$, respectively, which are almost the same but slightly slower than the simulated group velocity of $0.54c$ at 5.9 GHz. In addition, the measured valley-index-flipping time for the zigzag and armchair termination are $t_{rzig} = 3.17\pm0.06$ ns and $t_{rarm} = 0.15\pm0.12$ ns, respectively, verifying that valley-flipping rate of the zigzag termination is much lower than that of the armchair termination because of the near-conservation of the valley DOF at the zigzag termination. Note that the valley-flipping time delayed at the zigzag termination is much longer than the time required for a photon propagating through a lattice constant ($\frac{a}{v_g} \approx$ 0.34 ns). As a result, the electromagnetic energy is strongly localized near the terminated mirror and a subwavelength topology-controlled photonic cavity with a high-quality factor of $Q_{zig} = \pi f t_{rzig} = 58$ at frequency $f = 5.9$ GHz can be achieved near the mirror surface. In contrast, for the armchair termination, the valley-flipping time is of the same order as the time required for a photon propagating through a lattice constant, thus the electromagnetic energy cannot be tightly confined near the armchair termination and the topology-controlled photonic cavity exhibits a low-quality factor of $Q_{arm} = \pi f t_{rarm} = 2.78$ at frequency $f = 5.9$ GHz.

In conclusion, we have experimentally realized a novel type of topology-controlled photonic cavities by terminating a VPC waveguide with a judiciously oriented metallic mirror based on the near-conservation of the valley DOF. We experimentally observed that both the electromagnetic energy confinement and the quality factors of the topology-controlled photonic cavities are determined by the geometry of the metallic mirror, and the zigzag mirror exhibits a longer valley-index-flipping time than that of the armchair mirror because of the valley conservation. These results verify a new physical mechanism to localize electromagnetic waves and open a new avenue for designing topological photonic cavities with tunable field enhancement and quality factors.

Z.G. acknowledges the finding from the National Natural Science Foundation of China under grants No. 62361166627, 62375118, and 12104211, Guangdong Basic and Applied Basic Research Foundation under grant No.2024A1515012770, Shenzhen Science and Technology Innovation Commission under grants No. 20220815111105001 and 202308073000209,

High level of special funds under grant No. G03034K004. Y. M acknowledges the support from the National Natural Science Foundation of China under Grant No. 12304484, and the Guangdong Basic and Applied Basic Research Foundation under grant No. 2024A1515011371. B.Z. acknowledges the support from the National Research Foundation Singapore Competitive Research Program No. NRF-CRP23-2019-0007, and Singapore Ministry of Education Academic Research Fund Tier 2 under grant No. MOE2019-T2-2-085. The work at Sun Yat-sen University was sponsored by National Natural Science Foundation of China under Grant No. 12074443 and 12374364. The work at Jinan University was sponsored by the Natural Science Foundation of Guangdong Province General Program under Grant 2021A1515011944.

#These authors contributed equally
*chenxd67@mail.sysu.edu.cn
†guigeng001@e.ntu.edu.sg
‡blzhang@ntu.edu.sg
§gaoz@sustech.edu.cn